\documentstyle[epsfig,aps,prl]{revtex}
\begin{document}
\draft
\title{Nuclear modification of heavy quark fragmentation function and
$J/\psi$ production in ultrarelativistic heavy ion collisions}
\author{Bin Zhang, Bao-An Li, Andrew T. Sustich, and Charles Teal}
\address{Department of Chemistry and Physics,
Arkansas State University,\\
P.O. Box 419, State University,
Arkansas 72467-0419, USA}
\date{May 22, 2002}
\maketitle

\begin{abstract}
In ultrarelativistic heavy ion collisions, charm quark fragmentation 
is modified due to the melting of strings inside the high density
partonic matter. $D$ mesons produced in hadronization can further 
produce $J/\psi$ particles. Using a multiphase transport model, we 
investigate the effect on the rapidity distribution 
of the produced $J/\psi$ particles with two different charm quark
fragmentation functions. It is shown that the $J/\psi$ rapidity
distribution is sensitive to the nuclear modification of charm
quark fragmentation and thus is a good indicator of the onset
of string melting and the production of deconfined partonic matter.
\end{abstract}

One interesting topic in strong interaction physics 
that has attracted much attention is the study of the
phase transition between hot and dense hadronic matter and the Quark-Gluon
Plasma (QGP) \cite{fwilczek1}. The QGP is
believed to have existed in the early universe. It may also
exist in the cores of neutron stars. Knowledge
about the QGP phase transition is critical to the
understanding of strong interaction physics and several
other processes in cosmology and astrophysics.
Several accelerators have been built and many experimental
observables have been proposed for the search of the
QGP \cite{bmuller1,mgyulassy2}. 
$J/\psi$ production is one of the most promising probes for the
formation of 
the QGP \cite{tmatsui1,eshuryak1,xxu1,sgavin1,%
dkharzeev1,acapella1,cspieles1,wcassing1,%
dkahana1,bsa1,ko1,rvogt1,wcassing2,bzhang3,%
cgale1,rthews1,pbraunmunzinger2,lgrandchamp1,bzhang4}. 
In this Letter, we demonstrate that due to the melting
of strings inside a QGP, the heavy quark 
fragmentation function is modified. This modification
is reflected clearly in the $J/\psi$ rapidity distribution. The measurement
of the latter thus can be used to study
the onset of string melting and the production of the 
QGP.

In relativistic nuclear collisions, the charm quark distributions
can be calculated from perturbative quantum chromodynamics (pQCD). 
After the parton evolution, these charm quarks are converted into 
$D$ mesons. The $D$ meson distributions are determined by the 
charm quark distributions and the fragmentation
function. In hadron induced reactions, charm quark fragmentation
can be described by the delta function fragmentation function
\cite{hardprobe1},
\begin{equation}
D_c^D(z)=\delta(z-1).
\end{equation}
Charm quark fragmentation in $e^+e^-$ annihilation can be well
described by the Peterson fragmentation function \cite{peterson1},
\begin{equation}
D_c^D(z)=\frac{N}{z(1-1/z-\epsilon_c/(1-z))^2}.
\end{equation}
In the above equations, $z$ is the fractional momentum of the $D$
meson relative to the charm quark. $N$ is a normalization factor, and
$\epsilon_c\sim (m_q/m_c)^2$.
An explanation of the observed harder $D$ meson 
distribution in hadron induced interactions is provided by the
Lund fragmentation model. In the Lund model \cite{bandersson1} 
based event generator,
{\sc PYTHIA} \cite{tsjostrand1}, the charm quark is always at the 
end of a string and can
be pulled by the faster valence particles to acquire a harder distribution.

\begin{figure}[htb]
\centerline{\epsfig{file=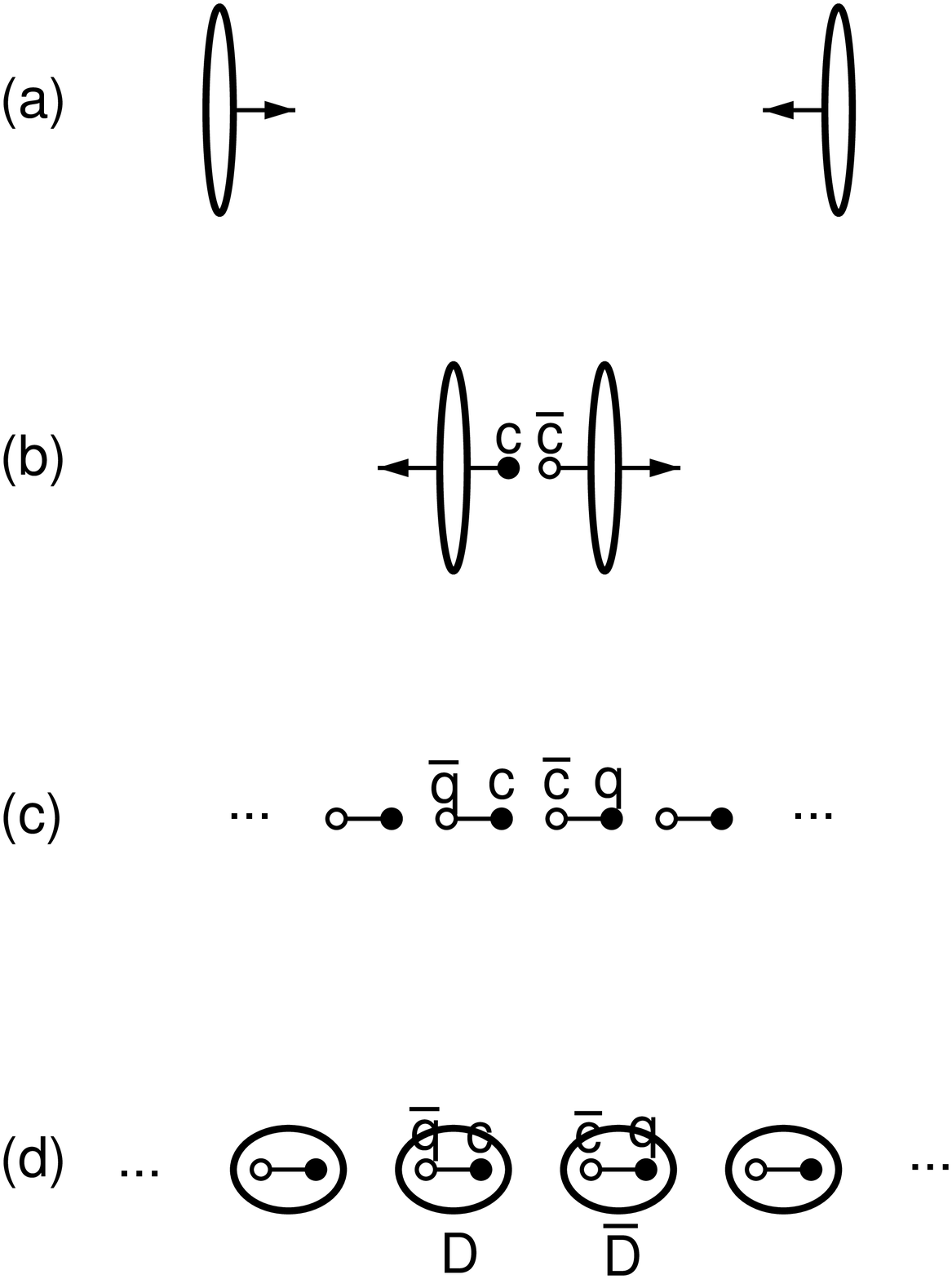,width=3.2in,height=3.6in,angle=0}}
\caption{
Time evolution of the collision in fragmentation scenario 1. 
(a) Two heavy nuclei at relativistic speeds approach each other. 
(b) After the collision, a charm quark and an anticharm quark 
are produced. They are attached to the receding nuclei by strings. 
(c) At a later time, quark-antiquark pairs are produced from the strings. 
(d) The charm quark and a light antiquark combine to form a $D$ meson, 
while the anticharm quark and a light quark combine to form a 
$\bar{D}$ meson.
}
\label{fig1}
\end{figure}

\begin{figure}[htb]
\centerline{\epsfig{file=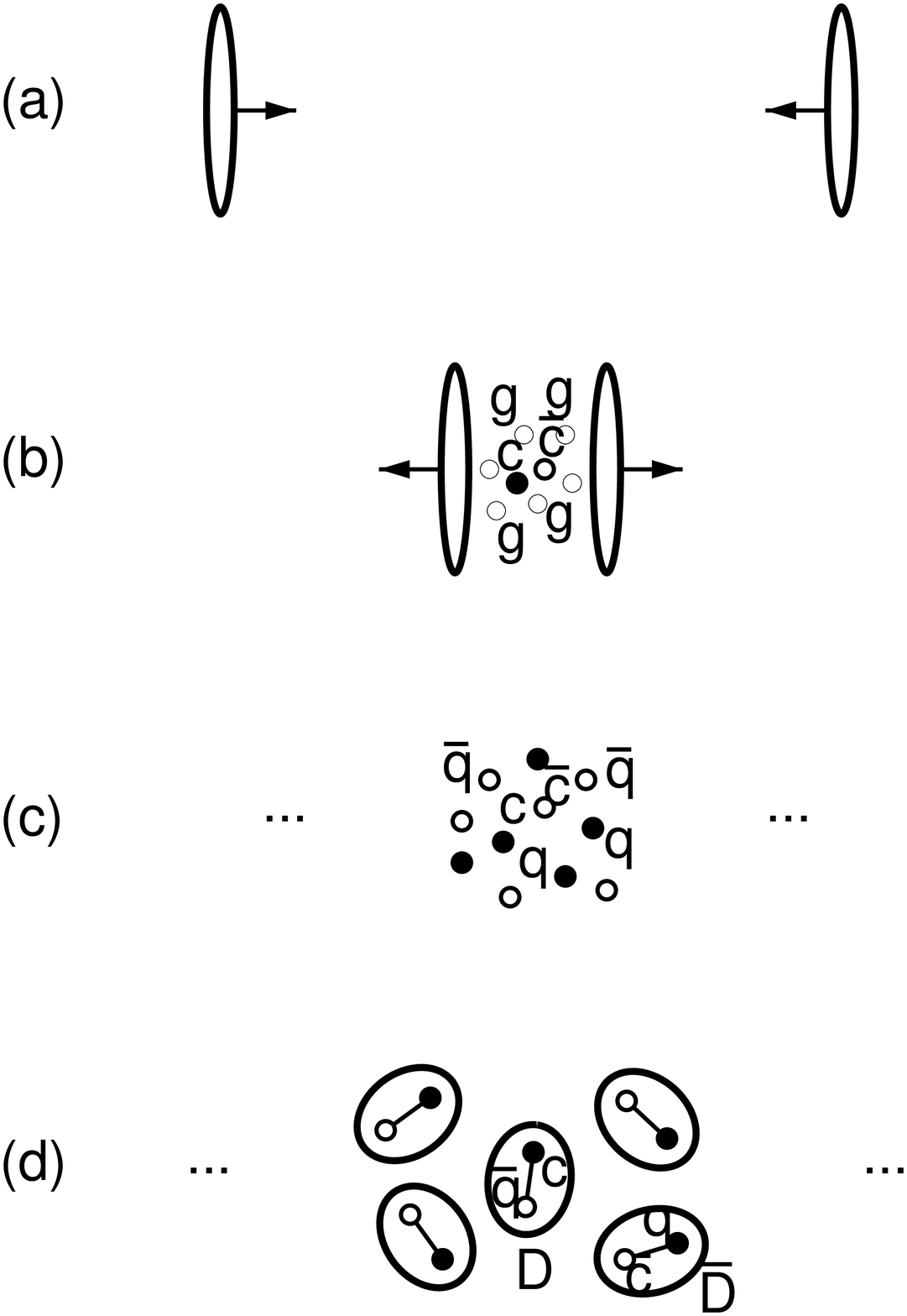,width=3.2in,height=3.6in,angle=0}}
\caption{
Time evolution of the collision in fragmentation scenario 2.
(a) Two relativistic nuclei approach each other.
(b) A charm quark and an anticharm quark are produced together with
a large number of gluons.
(c) At a later time, light quark-antiquark pairs are produced.
(d) The charm quark and a light antiquark combine to form a $D$ meson, 
while the anticharm quark and a light quark combine to form a 
$\bar{D}$ meson
}
\label{fig2}
\end{figure}

In nuclear collisions, two scenarios can happen. In scenario 1, 
the produced charm ($c$) and anticharm ($\bar{c}$) quarks are 
pulled by the strings of
the receding nuclei. These strings break up after being stretched. A
$c$ quark then combines with a light antiquark ($\bar{q}$) to form a $D$ meson.
A similar process happens for the formation of the $\bar{D}$ 
meson (Figure~\ref{fig1}). In scenario 2,
the $c$ and $\bar{c}$ quarks are produced
together with many gluons ($g$). The interactions between the charm quarks
and the receding nuclei are screened, and the strings between them
melt \cite{biro1,zlin3}. 
At a later time, gluons fragment into quark-antiquark pairs, while
other quark-antiquark pairs are also produced by the wounded nucleons.
A $D$ meson can be formed by the combination of a charm quark and a
light antiquark (Figure~\ref{fig2}).

These two scenarios can lead to very different rapidity distributions
for the produced $D$ and $\bar{D}$ mesons. In scenario 1, the charm
quarks are pulled by the faster receding hadrons. The rapidity
distribution will be harder and wider compared with that of scenario 2
in which the strings that pull the charm
quarks are absent. In the second scenario, the charm quarks are also 
slowed down by other quarks, and
the rapidity distribution will be softer and narrower.
Charm quark fragmentation in scenario 1 can be modeled by the
delta function fragmentation while that in scenario 2 can be modeled
by the Peterson fragmentation function. 

In ultrarelativistic nuclear collisions, the initial stage is dominated
by minijet gluons. In particular, for Au+Au collisions at 
$200A$ GeV, 
the pQCD calculated gluon rapidity density 
is about $300$ at midrapidity \cite{xnwang97a}.
For comparison, the charm and anticharm rapidity density
is about $3$ \cite{xnwang97a}. In other words, these charm and anticharm
quarks are immersed in a ``sea" of hot gluons. Unlike for
nuclear collisions at lower energies, or hadron-hadron collisions, 
the large amount of gluons modify the interactions between the charm
quarks and the valence quarks. The strings between them
melt when the gluon density is high. Several calculations have
demonstrated the possibility of string melting at the Relativistic
Heavy Ion Collider (RHIC) energies \cite{biro1,zlin3}.
The melting of strings then leads to the absence of the
pulling force that is responsible for the harder $D$ meson
distribution observed in hadron induced reactions. This modification
of the fragmentation function of charm quarks therefore reflects the 
production mechanism of the parton matter in
ultrarelativistic nuclear collisions.

We propose to use the $J/\psi$ rapidity distribution as a measure 
of this modification. $J/\psi$ production is affected by 
several factors.  Due to color
screening inside the QGP, $J/\psi$ particles
are dissociated above a certain temperature. A $J/\psi$ can be produced by
the collision of a $c$ quark and a $\bar{c}$ quark 
when the temperature is below the dissociation temperature. In
the parton phase, a $J/\psi$ can be destroyed by its collision with
a minijet gluon. Similar processes can occur in the hadron phase.
A $J/\psi$ can be produced by the collision of a $D$ meson and a $\bar{D}$
meson. It can be dissociated by the collision with a meson that is made of
light quarks.  
The above $J/\psi$ production from final state hadronic interactions 
relates the $J/\psi$ rapidity distribution
closely to the $D$ and $\bar{D}$ meson distributions. 

The modification of charm quark fragmentation and its 
reflection in the $J/\psi$ rapidity distribution are
studied by using a multiphase transport model \cite{bzhang1,zlin1,zlin2}. 
This model is a hybrid model that is made up
of several Monte Carlo models. The heavy ion jet interaction generator
({\sc HIJING}) \cite{xwang1,mgyulassy1} is used to simulate the 
passing of two heavy nuclei
and the production of the initial parton system. Zhang's parton
cascade ({\sc ZPC}) \cite{bzhang2} is then used to study the 
evolution of this parton
system. The conversion of the parton matter into hadron matter when
no more parton interactions are available is done using the 
Lund fragmentation \cite{bandersson1} that is part of 
the {\sc PYTHIA/JETSET} Monte Carlo package \cite{tsjostrand1}. 
A relativistic transport ({\sc ART}) \cite{bli1,bli2} 
model will follow the
hadron evolution until freeze-out. The multiphase model incorporates the
major channels in different stages. It has been used to study
particle production at RHIC energies and can successfully 
describe the global feature of the newly available RHIC data.

For $J/\psi$ production, $c+\bar{c}\leftrightarrow J/\psi+g$ is 
included in
the parton stage and $D+\bar{D}\leftrightarrow J/\psi+M$ 
is included in the
hadron stage. Here $M$ is a meson that is composed
of one light quark ($q$) and one light antiquark ($\bar{q}$). 
In relativistic nuclear collisions, charm quarks are 
rare particles. To effectively simulate the evolution of charm
particles and $J/\psi$'s, the perturbative 
method \cite{jrandrup1,xfang1} is used. 
In this
approach, charm quarks from several independent events are allowed
to evolve with a common hadron environment. Since charm quarks are 
rare particles, their effects on the common hadronic
environment are neglected.

We will compare the $J/\psi$ rapidity distributions in relativistic nuclear
collisions for the two fragmentation scenarios. The multiphase
transport model is used to simulate central Au+Au collisions at RHIC at
a center of mass energy, $\sqrt{s}=200A$ GeV. The {\sc PYTHIA} model is
used to generate the initial charm and anticharm quarks. 
The charm production cross section per nucleon-nucleon collision
is chosen to be 350 $\mu$b.
The mass of the charm quark is chosen to be 1.50 GeV, and the mass
of the $D$ mesons is set to be 2.01 GeV. A 3 mb cross section 
is used for the elastic 
collisions between gluons and charm quarks and
for the $J/\psi$ dissociation by gluons and light mesons. The 
cross sections of the inverse processes are calculated by using 
the detailed balance relation. These cross sections are consistent
with previous pQCD and hadron model calculations \cite{hardprobe1,%
combridge79a,peskin79a,khaglin1,zlin4,asibirtsev1,cywong1}.

\begin{figure}[htb]
\centerline{\epsfig{file=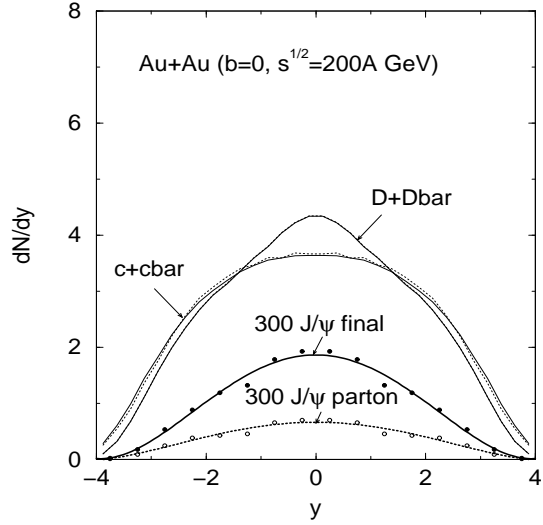,width=2.8in,height=2.8in,angle=-90}}
\null\vspace{0.5cm}
\caption{
Rapidity distributions of $J/\psi$ particles, charm and anticharm quarks,
$D$ and $\bar{D}$ mesons for the delta function fragmentation case.
The dotted line $J/\psi$ distribution is 300 times the distribution
after the parton stage. The solid line $J/\psi$ distribution is 300
times the distribution after the hadron evolution. The dotted line
charm and anticharm quark distribution is before the parton evolution
and the solid line is after the parton evolution.
The dotted line for the $D$ and $\bar{D}$
meson distribution is before the hadron evolution and the solid line is after
the hadron evolution.
}
\label{fig3}
\end{figure}

The rapidity distributions of the $J/\psi$ particle, together with those
for the charm quarks, $D$ and $\bar{D}$ mesons are calculated. Figure~\ref{fig3}
shows the results for the case in which the delta function fragmentation
is used to turn the charm quarks into $D$ mesons. It is seen that initial charm 
quark rapidity distribution has a plateau with a value of 
about 3.6 around
$y=0$. After the parton evolution, the final charm quark 
rapidity distribution is almost the same as the initial distribution. 
This distribution leads to a $J/\psi$ rapidity distribution
that also has a plateau with a value of about 0.7/300 around $y=0$.
The rapidity distribution for the $D$ and $\bar{D}$ mesons 
produced from hadronization is similar to the charm and anticharm quark
distribution.
The value around $y=0$ is about 4.3 and remains almost unchanged during the
hadron evolution. $J/\psi$ particles are produced further in the hadron stage
and the final rapidity distribution follows that of the $D$ and $\bar{D}$
mesons. The final value of $J/\psi$ particle around $y=0$ is about 
1.9/300.

\begin{figure}[htb]
\null\vspace{1cm}
\centerline{\epsfig{file=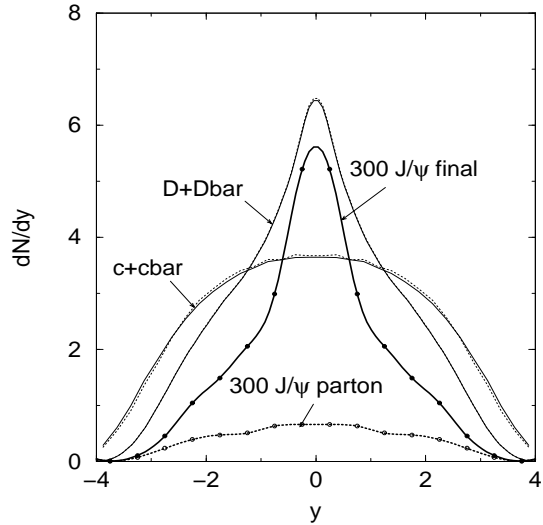,width=2.8in,height=2.8in,angle=-90}}
\null\vspace{0.5cm}
\caption{
Rapidity distributions of $J/\psi$ particles, charm and anticharm quarks,
$D$ and $\bar{D}$ mesons for the Peterson fragmentation function case.
The symbols are defined the same as those in Fig.~\ref{fig3}.
}
\label{fig4}
\end{figure}

The distributions for the Peterson fragmentation case are shown in 
Figure~\ref{fig4}. The initial, final charm quark distributions and 
the $J/\psi$
distribution after the parton stage are the same as 
those for the previous case since the parton stage
has no information about the hadronization. As expected, the $D$
and $\bar{D}$ rapidity distribution becomes narrower by
about 0.6 units of rapidity and has a
high peak with a value about 6.4 at $y=0$. Notice that the total number of
charm quarks does not change during hadronization. Hence, the area under
the rapidity distribution of the charm and anticharm quarks equals that of
the $D$ and $\bar{D}$ mesons. However, the production of $J/\psi$
particles is now starting from a $D$ and $\bar{D}$ rapidity distribution
that is different from that of the delta function fragmentation case. This
leads to a final $J/\psi$ distribution that is also narrower and has a 
peak in the middle with a value of about 5.6/300 around $y=0$.

In conclusion, the production of deconfined matter 
modifies the fragmentation of the charm and anticharm quarks.
In nuclear collisions at lower energies, the string dynamics is important,
and the charm fragmentation is similar to that in the hadron collisions.
In nuclear collisions at ultrarelativistic energies, 
the minijet dynamics becomes important,
and the charm fragmentation is more like the independent fragmentation
in electron-positron collisions. This modification leads
to a rapidity distribution of $J/\psi$ distinctively 
different from what is normally 
expected in hadronic reactions.  The $J/\psi$ rapidity distribution
measurement, in conjunction with information about charmed mesons,
can offer valuable insight about the melting of strings and the
production of the QGP in relativistic nuclear collisions at RHIC.

This work was supported in part by the Arkansas Science and Technology 
Authority under Grant No. 01-B-20, and by the U.S. National Science 
Foundation under Grant No. 0088934. We thank Z.W. Lin for helpful
discussions. We also thank the Parallel Distributed
System Facility at the National Energy Research Scientific Computer Center
for providing computer resources. B.A. Li would like to thank S. Das
Gupta and C. Gale for their kind hospitality at McGill University
where part of this work was done.

\clearpage

\end{document}